\documentclass[twocolumn,showpacs,preprintnumbers,amsmath,amssymb,aps,prl,reprint]{revtex4}
\usepackage{graphicx}
\begin{document}

\title{
Shapiro Spikes and Negative Mobility for Skyrmion Motion on Quasi-One Dimensional Periodic Substrates  
} 
\author{
C. Reichhardt and C. J. Olson Reichhardt 
} 
\affiliation{
Theoretical Division and Center for Nonlinear Studies,
Los Alamos National Laboratory, Los Alamos, New Mexico 87545, USA 
} 

\date{\today}
\begin{abstract}
  Using a simple numerical model of skyrmions in a two-dimensional system interacting with
  a quasi-one dimensional
  periodic substrate under combined dc and ac drives where the dc drive is
  applied perpendicular to the substrate periodicity,
  we show that a rich variety of novel phase locking dynamics
  can occur due to the influence of the Magnus
  term on the skyrmion dynamics.
  Instead of Shapiro steps,
 the velocity response in the direction of the dc drive exhibits
  a series of spikes,
  including 
  extended dc drive intervals over which the skyrmions move in the direction
  opposite to the dc drive, producing negative mobility.
  There are also specific dc drive values at which the skyrmions move
  exactly perpendicular to the dc drive direction,
  giving a condition of absolute transverse mobility.  
\end{abstract}
\maketitle

When an overdamped particle is driven by a combined dc and ac drive over a
periodic substrate,
a series of steps, called Shapiro steps \cite{1}, appear in the velocity response
over fixed dc drive intervals
due to phase locking between the ac driving frequency and the
oscillatory frequency of the particle motion induced by the substrate periodicity.
Phase locking of this type occurs for dc plus ac driven
Josephson junction arrays \cite{2}, sliding
charge density waves \cite{3}, vortices in type-II superconductors \cite{4,5,6}
or colloids \cite{7} moving over periodic pinning
arrays, frictional systems \cite{8}, and numerous other nonlinear systems
in which there are two coupled competing frequencies \cite{9,10,11}.
In a two dimensional (2D) overdamped system
with a quasi-one-dimensional (q1D) substrate,
Shapiro steps only occur when the dc and ac drives are both applied
parallel to the substrate periodicity direction,
since the pinning does not induce a periodic modulation of the particle motion 
for perpendicular driving.
In some systems, additional non-dissipative terms can be relevant to the
particle dynamics,
such as a Magnus force
which generates a particle velocity component that is
perpendicular to the net applied force on the particle.
Magnus effects are known to be important for skyrmions
in chiral magnets,
where the ratio of the Magnus term to the damping term
can be ten or higher \cite{12,13,14,15,16,17}.
Skyrmions can be set into motion by an applied
spin-polarized current, and the Magnus term has been shown to strongly affect the
interaction of the moving skyrmions with pinning sites,
leading to reduced depinning thresholds \cite{14,15,16,18,19},
a drive dependent skyrmion Hall angle \cite{19,20,21,22},
and skyrmion speed up effects \cite{20,21}.

Recent studies of
skyrmions driven over a periodic q1D substrate by a dc drive that is parallel to
the substrate periodicity direction combined with a perpendicular ac drive
showed that
a new class of Magnus-induced Shapiro steps arises due to
an effective coupling by the Magnus term of the
perpendicular and parallel particle motion, whereas in the
overdamped limit no Shapiro steps occur for this drive configuration
\cite{23}.
Here we examine
skyrmions confined to a 2D plane containing a q1D periodic substrate
and moving under the influence of a dc drive applied
{\it perpendicular} to the substrate
periodicity direction along with a parallel or perpendicular ac drive,
and we find that a rich variety of
dynamical phases can occur.
Instead of Shapiro steps,
the particle velocity response in the dc drive direction
exhibits what
we call Shapiro spikes where the slope of the velocity-force curve
locks to a constant value
over a range of dc driving forces.
One of the most remarkable features of this system
is that there are also a series of extended dc drive regions where
the particle motion is in the direction
{\it opposite} to the dc drive,  known as negative mobility \cite{24,25,26}.
It is even possible for the particle motion
at some drives to be exactly perpendicular
to the dc drive direction, creating a condition of absolute transverse mobility \cite{27}.
Negative mobility effects
have been observed in overdamped systems but generally require more
complicated substrates, thermal fluctuations, many-particle collective effects, or
the application of multiple ac drives, whereas
in the skyrmion system, negative mobility arises for a much simpler set of conditions.
We map the evolution of the dynamic phases as a function of ac
drive amplitude and the ratio of the Magnus to the damping term.
In addition to their interest as signatures of a new dynamical system,
these results could be important
in providing a new way to precisely control the direction of motion of skyrmions
in order to realize skyrmion-based memory or logic devices \cite{28}.

{\it Simulation-- } 
We model a 2D system with periodic boundary
conditions in the $x$ and $y$ directions
containing a q1D substrate and a skyrmion treated with a particle-based
model that has previously been used to
examine driven skyrmion motion in random\cite{18,19},
2D periodic \cite{21}, and 1D periodic substrates \cite{23,29}.
The skyrmion dynamics are determined using
the following equation of motion:
\begin{equation}
\alpha_d {\bf v}_{i} + \alpha_m {\hat z} \times {\bf v}_{i} =
{\bf F}^{sp}_{i} + {\bf F}_{dc} + {\bf F}_{ac}  ,
\end{equation}
where the skyrmion velocity is
${\bf v}_{i} = {d {\bf r}_{i}}/{dt}$.
On the left hand side, $\alpha_d$ gives the strength of the damping term,
which aligns the skyrmion velocity in the direction of the net external forces,
while $\alpha_m$ is the Magnus term, which
rotates the velocity in the direction perpendicular to
the net external forces.
For varied ratios of $\alpha_{m}/\alpha_{d}$
we impose the constraint $\alpha_d^2 + \alpha_m^2 = 1$.
The force from the substrate is ${\bf F}^{sp}_i = \nabla U(x_i){\hat {\bf x}}$
where $U(x) = U_{o}\cos (2\pi x/a)$
and $a$ is the substrate lattice constant.
  The substrate strength
is defined to be $A_{p} \equiv 2\pi U_{0}/a$.
The dc drive ${\bf F}_{dc}=F_{dc}^{\perp}{\bf \hat y}$ is applied perpendicular to the
substrate periodicity direction, while the
ac driving force ${\bf F}_{ac}=F_{ac}^{||}{\bf \hat x}$
or ${\bf F}_{ac}=F_{ac}^{\perp}{\bf \hat y}$ is applied either
parallel or perpendicular to the substrate periodicity direction, respectively.
We characterize  the system by measuring the velocity response
$V_{||}=2\pi \langle V_x\rangle /\omega a$ parallel to the substrate periodicity and
$V_{\perp}=2\pi \langle V_y\rangle /\omega a$ perpendicular to the substrate periodicity,
so that on a Shapiro
step the velocity is integer valued with $V_{||}=n$ or $V_{\perp}=n$, allowing us to
identify the step number $n$.

\begin{figure}
\includegraphics[width=\columnwidth]{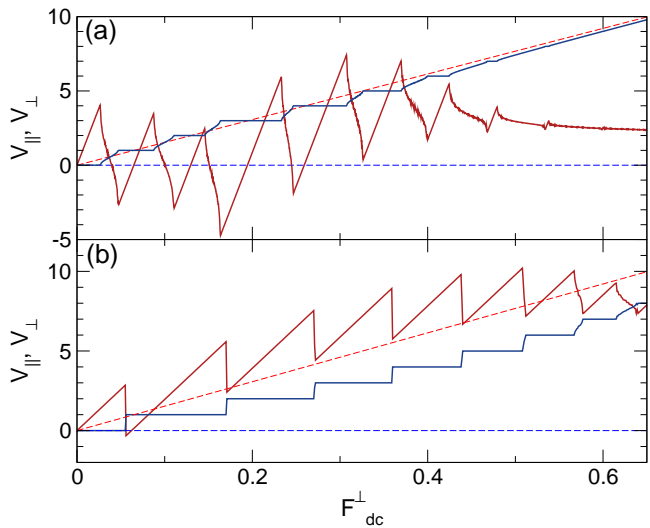}
\caption{(a) The average skyrmion velocity $V_{||}$ (blue) and
  $V_{\perp}$ (red) vs $F^{\perp}_{dc}$
  for a system with
pinning strength $A_{p} = 1.0$
and perpendicular ac drive $F^{\perp}_{ac} = 0.325$
at different
ratios $\alpha_m/\alpha_d$ 
of the Magnus to dissipative terms.  Solid lines:
$\alpha_m/\alpha_d = 9.96$; dashed lines:
$\alpha_m/\alpha_d=0$.
For $\alpha_m/\alpha_d=9.96$, there are Shapiro steps in
$V_{||}$
and spikes in $V_{\perp}$, along with intervals in which
$V_{\perp}< 0$. 
(b) The same for
$\alpha_m/\alpha_d=3.219$ (solid lines) and
$\alpha_m/\alpha_d=0$ (dashed lines).  
}
\label{fig:1}
\end{figure}

{\it Results and Discussion---}
In Fig.~\ref{fig:1}(a) we plot $V_{||}$ and $V_{\perp}$ versus $F^{\perp}_{dc}$
for a system with $A_{p} = 1.0$, $\alpha_{m}/\alpha_{d} = 9.96$,
and $F^{\perp}_{ac} = 0.325$.
The dashed lines show the average velocities in
the overdamped limit of $\alpha_{m}/\alpha_{d} = 0$,
where the particles simply slide along the $y$-direction with an Ohmic
response and phase locking does not occur.
When the Magnus term is finite,
$V_{||}$ shows a series of phase-locked Shapiro steps,
while $V_{\perp}$ shows a completely different response consisting of
spike like features.  On each phase-locked step in $V_{||}$, the slope
of $V_{\perp}$ is constant.
The most remarkable feature in $V_{\perp}$ is that there
are four extended intervals of $F^{\perp}_{dc}$
over which  $V_{\perp}< 0$, indicating that the particle is moving
in the {\it opposite} direction to the applied dc drive,
a phenomenon known as negative mobility \cite{25,26}.
On a given step,
$V_{\perp}$ can grow from negative values to positive values, passing through a point
at which
$V_{||}$ is finite but $V_{\perp} = 0$,
indicating that particle is moving exactly perpendicular to the
applied dc drive in a phenomenon known as
absolute transverse mobility \cite{27}.
At higher values of $F^{\perp}_{dc}$, the negative mobility regions
are lost and the minimum value of $V_{\perp}$ at the bottom of each spike
increases with increasing $F^{\perp}_{dc}$.
At the top of the $V_{\perp}$ spikes,
the particle velocity in the dc drive direction
is higher than it would be in an overdamped system, which
is an example of a pinning-induced speed up effect \cite{20,21}.
After each spike, $V_{\perp}$ decreases with
with increasing $F^{\perp}_{dc}$, which is an example of negative differential conductivity.
In Fig.~\ref{fig:1}(b) we show that for
$\alpha_{m}/\alpha_{d} = 3.219$,
there are still spikes in $V_{\perp}$; however, the regions of negative mobility are lost.
As $\alpha_{m}/\alpha_{d}$ is further reduced,
$V_{\perp}$ gradually becomes smoother and approaches the dashed line, which
indicates the response in the overdamped limit.

\begin{figure}
\includegraphics[width=\columnwidth]{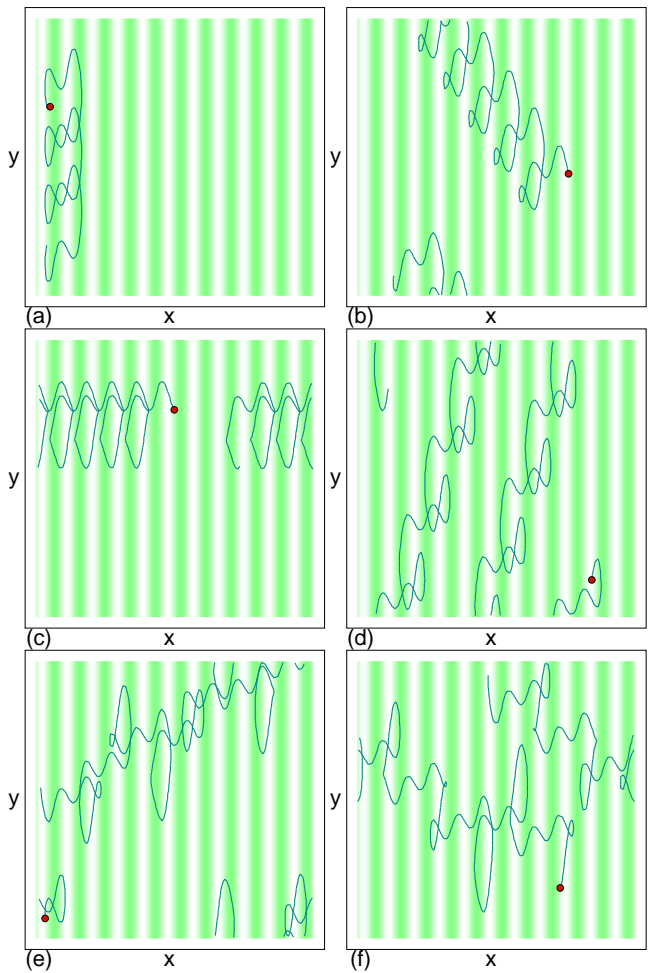}
\caption{Skyrmion location (dot) and trajectory (line) on a
  q1D periodic substrate potential for
  ac and dc drives both applied along the
  perpendicular or $y$ direction for the system in Fig.~\ref{fig:1}(a) with
  $\alpha_m/\alpha_d=9.96$.
  The lighter regions indicate the locations of the
  substrate minima.
  (a) $F^{\perp}_{dc} = 0.015$ along the $n = 0$ step.
  (b) $F^{\perp}_{dc} = 0.055$, showing a phase locked region
with negative mobility where $V_{\perp} < 0$ and $V_{||} > 0$.   
(c) $F^{\perp}_{dc} = 0.064815$, where there is absolute transverse mobility
with $V_{\perp} = 0$ and $V_{||} > 0.$ 
(d) $F^{\perp}_{dc} = 0.085$ along the $n = 1$ step, where
$V_{\perp}$ and $V_{||}$ are both positive. 
(e) $F^{\perp}_{dc} = 0.095$, where there is a non-phase locked region
with  $V_{\perp} > 0$ 
and $V_{||} > 0$.
(f) $F^{\perp}_{dc}=0.105$, where there is a non-phase locked region with negative mobility. 
}
\label{fig:2}
\end{figure}

From the dynamics in Fig.~\ref{fig:1}(a) we define six
different regimes for the particle motion.
Region I is a phase locked
state in which the particle moves in a periodic orbit with $V_{\perp} > 0$ and
$V_{||} \geq 0$.
In Fig.~\ref{fig:2}(a) we show the particle trajectory
at $F^{\perp}_{dc} = 0.015$, corresponding to the $n = 0$
step where the particle orbit translates only along the $y$
direction.
Figure~\ref{fig:2}(d) illustrates the $n=1$ step at
$F^{\perp}_{dc} = 0.085$, where the particle
moves in a periodic orbit that translates in both the positive $x$ and $y$ directions.
Region II is a phase locked state in which
the particle moves in the direction opposite to the dc driving force with
$V_{\perp} < 0$,
as shown in Fig.~\ref{fig:2}(b) for $F^{\perp}_{dc} = 0.055$
on the $n= 1$ step where
the periodic particle orbit translates in the positive $x$ and
negative $y$ directions.
In Region III, which is also phase locked,
the particle exhibits absolute transverse mobility and
moves strictly in the positive $x$-direction
with $V_{\perp} = 0$, as shown in Fig.~\ref{fig:2}(c)
at $F^{\perp}_{dc} = 0.064815$.
This corresponds to a skyrmion Hall angle of $\theta_{sk}=90^{\circ}$.
Region IV is a non-phase locked state in which
$V_{\perp} = 0$ while $V_{||}$ is positive.
It occurs in the non-step regions where the particle does
not follow a periodic orbit and does not translate along the $y$
direction,
such as near $F^{\perp}_{dc} = 0.04$ in Fig.~\ref{fig:1}(a).
The absolute transverse mobility of Regions III and IV only occurs at
specific values of $F^{\perp}_{dc}$,
while the other phases span extended intervals of the dc driving force.
Region V is a
non-phase locked state where $V_{\perp}$ and $V_{||}$ are both positive
but the particle does not form a periodic orbit, as illustrated in
Fig.~\ref{fig:2}(e) at $F^{\perp}_{dc} = 0.095$.
Finally, Region VI is a
non-phase locked state
in which $V_{\perp} < 0$ and $V_{||} > 0$,  as shown in
Fig.~\ref{fig:2}(f) at  $F^{\perp}_{dc} = 0.105$.
We note that there can be smaller intervals outside of the integer phase locked steps
over which the system can exhibit fractional phase locking, and that these fractional steps
become more prominent for
higher values of $F^{\perp}_{ac}$.

\begin{figure}
\includegraphics[width=\columnwidth]{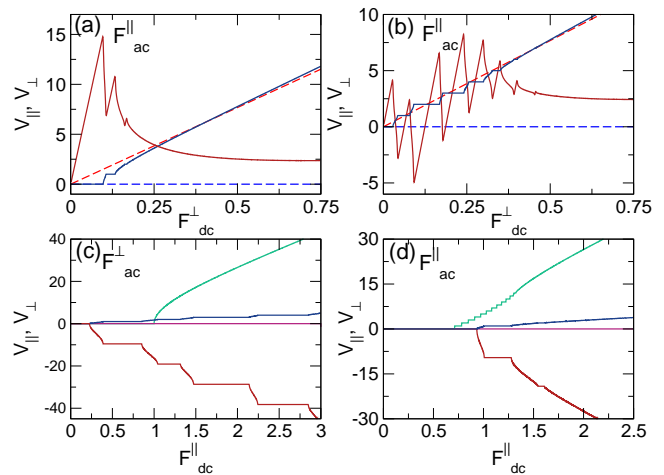}
\caption{
  (a) $V_{||}$ (blue) and $V_{\perp}$ (red) vs $F^{\perp}_{dc}$ for
  perpendicular dc driving
  and parallel ac driving
  at 
  $F^{||}_{ac} = 0.325$.
  Solid lines: $\alpha_m/\alpha_d=9.96$; dashed lines: $\alpha_m/\alpha_d=0$.
  (b) The same for $F^{||}_{ac} = 2.35$, where there
intervals in which $V_{\perp} < 0$. 
(c) $V_{||}$ (blue, green) and $V_{\perp}$ (red, purple) vs $F^{||}_{dc}$ for
parallel dc driving
and
perpendicular ac driving
at
$F^{\perp}_{ac} = 0.325$ 
for $\alpha_{m}/\alpha_{d} = 9.96$ (blue and red), showing Shapiro steps,
and for
$\alpha_{m}/\alpha_{d} = 0$ (green and purple), where no Shapiro steps occur.
(d) $V_{||}$ (blue, green) and $V_{\perp}$ (red, purple) vs $F^{||}_{dc}$ for
parallel dc driving
and parallel ac driving
at $F^{||}_{ac} = 0.325$
for $\alpha_{m}/\alpha_{d} = 9.96$ (blue and red)
and  $\alpha_{m}/\alpha_{d} = 0$ (green and purple), showing Shapiro steps.
}
\label{fig:3}
\end{figure}

We next consider the case
of a perpendicular dc drive and a parallel ac drive,
as shown in Fig.~\ref{fig:3}(a)
where we plot $V_\perp$ and $V_{||}$ vs $F^{\perp}_{dc}$ for a system with the same
parameters as in Fig.~\ref{fig:1}(a)
for
$F^{||}_{ac} = 0.325$. 
Here, for $\alpha_m/\alpha_d=9.96$,
there are still steps in $V_{||}$ and spikes in $V_{\perp}$; 
however, $V_{\perp} \geq 0$ for all $F^{\perp}_{dc}$.
For the overdamped $\alpha_m/\alpha_d=0$ case, $V_{||} = 0$ and 
$V_{\perp}$ increases linearly with increasing $F^{\perp}_{dc}$.
In Fig.~\ref{fig:3}(b),
for the same driving configuration at $F^{||}_{ac} = 2.35$,  
there are
more steps in $V_{\perp}$ as
well as regions in which $V_{\perp} < 0$,
similar to the perpendicular ac driving case in Fig.~\ref{fig:1}(a).
This shows that it is possible to observe negative mobility and spike features
in $V_{\perp}$ whenever the dc drive is applied perpendicular to the substrate periodicity,
regardless of the ac driving direction.
The ac drive amplitudes at which the features appear are much lower
for perpendicular ac driving than for parallel ac driving.

For comparison, Fig.~\ref{fig:3}(c) shows the results of applying a parallel
dc drive $F_d^{||}$
and
a perpendicular ac drive
with $F^{\perp}_{ac} = 0.325$ at $\alpha_{m}/\alpha_{d} = 9.96$.
Here, phase locking steps are present but the spikes associated with negative
mobility are not.  The curves for the overdamped case with $\alpha_m/\alpha_d=0$
show that $V_{||}$ has a finite depinning threshold but no Shapiro steps,
while $V_{\perp}=0$ for all $F^{||}_{dc}$.
In Fig.~\ref{fig:3}(d), both the ac and dc drives are parallel to the substrate periodicity
with $F^{||}_{ac} = 0.325$.
At $\alpha_{m}/\alpha_{d}  = 9.96$, both $V_{||}$ and $V_{\perp}$ exhibit
Shapiro steps, while
in the overdamped limit with $\alpha_m/\alpha_d=0$,
$V_{||}$ contains Shapiro steps while $V_{\perp} = 0.$
This shows that in the overdamped limit, Shapiro
steps occur only when both the ac and dc driving
are applied parallel to the substrate periodicity direction.

The negative mobility for perpendicular dc driving arises due to the combination of the
Magnus term and the skyrmion-pinning interactions. 
Under a finite Magnus term,
the dc drive generates
an $x$ direction force 
$F_x=F^{\perp}_{dc}\sin (\theta_{sk})$
on the skyrmion, where 
$\theta_{sk}=\tan^{-1}(\alpha_m/\alpha_d)$.
In response, the substrate exerts
an $x$ direction force on the skyrmion that
the Magnus term transforms
into a $y$ velocity component
in the range $V_{y} = \pm A_{p}\sin(\theta_{sk})$.
For certain intervals of
$F^{\perp}_{dc}$, the $-y$ portion of the ac driving cycle
synchronizes with the time at which the pinning force
generates a $-y$ velocity component,
resulting in a
net negative value of $V_{\perp}$.
Conversely, in other 
$F^{\perp}_{dc}$ intervals the $+y$ portion of the ac driving cycle
synchronizes with the time at which
the substrate generates a $+y$ velocity component,
producing a speed up effect with enhanced
positive $V_{\perp}$.
Somewhere between these two intervals, $V_{\perp} = 0$ and
absolute transverse mobility occurs.
All of these effects become stronger for
higher ac amplitude and larger ratios of $\alpha_{m}/\alpha_{d}$.
A similar argument can be made for ac driving in the $x$-direction; however,
the ac amplitude must be
larger by a factor of approximately $\alpha_{m}/\alpha_{d}$,
such as shown in Fig.~\ref{fig:3}(b), for effects of the same magnitude to
occur,
since for a parallel ac drive the $y$-velocity component is
multiplied by a factor of
$\cos(\theta_{sk})$ instead of $\sin(\theta_{sk})$.

\begin{figure}
\includegraphics[width=\columnwidth]{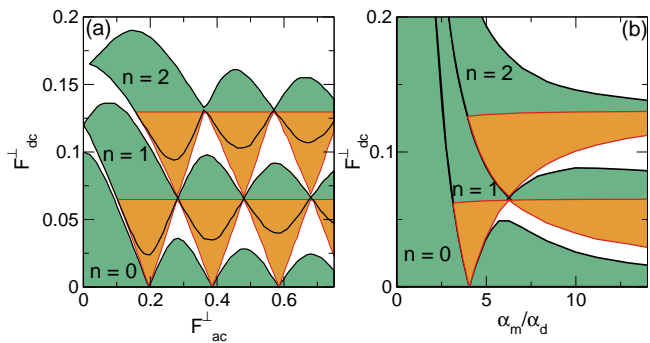}
\caption{
  (a) Dynamic phase diagram for $F^{\perp}_{dc}$
  vs $F^{\perp}_{ac}$ showing the locations of the $n = 0$, 1, and $2$ steps
  (outlined in black)
  for $\alpha_{m}/\alpha_{d} = 9.96$.
  Green: phase locked regions with $V_{\perp} > 0$;
  white: unlocked regions with $V_{\perp} > 0$;
  orange: locked or unlocked regions with $V_{\perp} < 0$.
  Along the red lines, $V_{\perp} = 0$ and $V_{||} > 0$.
  (b) Dynamic phase diagram for $F^{\perp}_{dc}$ vs
  $\alpha_{m}/\alpha_{d}$ at $F^{\perp}_{ac}=0.325$.  Colors are the same as in
  panel (a).
}
\label{fig:4}
\end{figure}

In Fig.~\ref{fig:4}(a) we plot the evolution of the different regimes
for the system in Fig.~\ref{fig:1}(a) as a function
of $F^{\perp}_{dc}$ and $F^{\perp}_{ac}$, focusing
only on the
$n = 0$, 1, and $2$ phase-locked regions.
The width of the $n$-th phase locked step has the same
$J_n$ or Bessel function
oscillating behavior predicted to occur for Shapiro steps 
\cite{30}.
The green shading denotes phase locked regimes
with $V_{\perp} > 0$.
White indicates unlocked regions with $V_{\perp} > 0$.
The orange shading indicates phase locked and unlocked
regions of negative mobility with $V_{\perp} < 0$, which
form a series of triangles
that overlap with the $n = 1$ and $2$ steps.
At the edges of these triangles, absolute transverse mobility with $V_{\perp} = 0$ and
$V_{||} > 0$ occurs.
We observe similar dynamic phases for steps with higher values of $n$.
This result indicates that the direction of the
skyrmion motion can be tuned by varying either the dc or ac perpendicular drives.
In Fig.~\ref{fig:4}(b) we plot a dynamic phase diagram as a function of
$F^{\perp}_{dc}$ and $\alpha_{m}/\alpha_{d}$ at $F^{\perp}_{ac} = 0.325$.
Here, 
for small $\alpha_{m}/\alpha_{d}$
the skyrmion motion is locked in the  perpendicular direction.
Negative mobility occurs
only for $\alpha_{m}/\alpha_{d} > 3.2$, and higher order steps
emerge as $\alpha_m/\alpha_d$ increases.
Similar phase diagrams can be created for 
parallel ac driving; however, in this case,
negative mobility does not occur until much higher ac driving amplitudes are
applied.
We also find that these effects are robust for multiple interacting skyrmions
when the skyrmion-skyrmion interactions are modeled as a repulsive force.

{\it Summary---} 
We have shown that when a skyrmion
obeying dynamics that are governed by both a Magnus and a dissipative term moves
under combined ac and dc drives on a
quasi-1D periodic substrate,
a rich variety of phase locking phenomena can occur
that
are absent in the overdamped limit.
When the dc drive is applied perpendicular to the substrate periodicity direction,
for either parallel or perpendicular ac driving
the perpendicular velocity response develops
Shapiro steps, while the parallel velocity exhibits
Shapiro spikes.
We also observe extended dc drive intervals
over which
the skyrmion moves in the opposite direction to the
dc drive, known as negative mobility,
while for specific dc drive values we find absolute transverse mobility in which the
skyrmion moves exactly transverse to the dc drive.
When the dc drive is applied parallel to the substrate periodicity direction,
the Shapiro spikes and negative mobility are absent,
while in the overdamped limit Shapiro steps only occur when  the dc and ac drives
are both applied parallel to the substrate periodicity direction.
The dynamics we observe
should be realizable for skyrmions in chiral magnets interacting with
quasi-1D substrates created using
1D thickness modulations or line pinning arrays,
and open a new way to control skyrmion motion.

\begin{acknowledgments}
We gratefully acknowledge the support of the U.S. Department of
Energy through the LANL/LDRD program for this work.
This work was carried out under the auspices of the 
NNSA of the 
U.S. DoE
at 
LANL
under Contract No.
DE-AC52-06NA25396 and through the LANL/LDRD program.
\end{acknowledgments}

\end{document}